\documentstyle[11pt, titlepage]{article}
\newcommand{\beq}{\begin{equation}}
\newcommand{\beqn}{\begin{eqnarray}}
\newcommand{\eeq}{\end{equation}}
\newcommand{\eeqn}{\end{eqnarray}}
\begin{document}
\begin{titlepage}
\rightline {Preprint HUTP-94/A011}
\rightline {astro-ph/9405029}
\begin{center}
\bigskip \bigskip \bigskip \bigskip \bigskip
\Large\bf Induced-gravity Inflation and the \\
Density Perturbation Spectrum\\
\bigskip\bigskip\bigskip
\normalsize\rm
David I. Kaiser \\
\bigskip
\it Lyman Laboratory of Physics\\
Harvard University\\
Cambridge MA 02138 \\ \rm
e-mail:  dkaiser@husc.harvard.edu\\
\bigskip\bigskip\bigskip\bigskip
27 July 1994\\ \bigskip
\bigskip
\bigskip\bigskip\bigskip\bigskip\bigskip\bigskip
\bf Abstract \rm
\end{center}
\narrower Recent experimental determinations of the
spectral index describing the scalar mode spectrum of
density perturbations encourage comparison with
predictions from models of the very early universe.
Unlike extended inflation, Induced-gravity Inflation
predicts a power spectrum with $0.98 \leq n_s \leq 1.00$,
in close agreement with the experimental measurements. \bigskip
\end{titlepage}
\newpage

%_________________________
\baselineskip 28pt
\indent  An exciting test for models of the very early
universe stems from recent measurements of the power
spectrum of density perturbations, as seen in the cosmic
microwave background radiation.  The scalar spectrum,
modeled as $\cal{P} \rm (\it k \rm) \propto \it k^{n_s}
\rm$~\cite{pr93}, functions as a test for models like inflation,
independently of the familiar test based on the
magnitude of the fluctuations.  As pointed out by Andrew
Liddle and David Lyth~\cite{lidlyth92}, extended inflation
predicts a spectral index ($n_s$) which is tilted too far
away from the Harrison-Zel'dovich (scale-invariant)
spectrum ($n_s = 1.00$), and hence cannot match the
recent Cosmic Background Explorer (COBE) determination.
In this Letter, the
predictions from a cousin-model of extended inflation,
Induced-gravity Inflation, are compared with the experimental
values.  Unlike extended inflation, Induced-gravity
Inflation predicts a spectral index in quite close
agreement with recent experimental values. \\
\indent  Like extended inflation~\cite{EI},
Induced-gravity Inflation (IgI)~\cite{AZT85}~\cite{DK94}~\cite{FakUn90}
incorporates a Generalized Einstein Theory (GET) gravity
sector.  Yet unlike extended inflation, IgI incorporates
only one scalar boson to get
all the work of inflation done:  the scalar field which
couples to the Ricci scalar in Brans-Dicke-like fashion is
the $same$ field whose potential, $V(\phi)$, drives
inflation.  This is the crucial difference as far as $n_s$ is
concerned:  by adopting a potential which leads to a
$second$ order phase transition (unlike the first order
phase transition incorporated in extended inflation), IgI
can escape two related problems of extended inflation
(discussed below) and lead to an acceptable spectrum of
perturbations. \\
\indent  Much of the formalism developed in the literature for
calculating
$n_s$ assumes an Einsteinian gravitational background~\cite{ns}; hence it
cannot be applied in a straightforward manner to IgI, because of its
GET gravity sector.  Furthermore, as discussed in~\cite{FakHabib}, the
usual strategy of applying a conformal transformation to bring IgI into the
canonical Einstein-Hilbert gravitational form~\cite{conftrans} may prove
problematic when studying the spectrum of perturbations, stemming from
ambiguities with semiclassical quantization in the various frames.
Therefore, in
this Letter, we will restrict attention to the \lq\lq physical" or \lq\lq
Jordan" frame, in which the nonminimal $\phi^2 R$ coupling is explicit. \\
\indent  The action for IgI is given by:
%%%%%%%%%%[Equation 1]%%%%%%%%%%
\beq
S = \int d^4 x \> \sqrt{-g} \> \left( \frac{\phi^2}{8\omega}
R - \frac{1}{2} \phi_{; \mu} \phi^{; \mu} - V(\phi) + L_M
\right) ,
\eeq
where the Brans-Dicke parameter ($\omega$) is related to the nonminimal
curvature coupling constant ($\xi$), often found in the literature, by
$\omega = (4 \xi)^{-1}$.  In this model, $L_M$ only includes contributions
from \lq ordinary' matter, and does not include a separate Higgs
sector; it can henceforth be ignored.  The coupled field
equations which result are:
%%%%%%%%%%[Equation 2]%%%%%%%%%%
\beqn
\nonumber H^2 + \frac{k}{a^2} = \frac{4\omega}{3
\phi^2} V (\phi) + \frac{2\omega}{3}
\left(\frac{\dot{\phi}}{\phi}\right)^2 - 2H \left(
\frac{\dot{\phi}}{\phi} \right) + \frac{2 (\omega +
1)}{3a^2} \frac{(\partial_i \phi)^2}{\phi^2} +
\frac{2}{3a^2} \frac{\partial_i^2 \phi}{\phi} , \\
\ddot{\phi} + 3H \dot{\phi} + \frac{\dot{\phi}^2}{\phi} -
\frac{1}{a^2} \frac{(\partial_i \phi)^2}{\phi} -\frac{1}{a^2}
\partial_i^2 \phi = \frac{2\omega}{3 + 2\omega} \phi^{-1}
\left[ 4 V(\phi) - \phi V^{\prime} (\phi) \right] .
\eeqn
In equation (2), $a(t)$ is the cosmic scale factor of the
Robertson-Walker metric, and is related to the Hubble
parameter by $H \equiv \dot{a} / {a}$.  Note from the
form of equation (1) that $4\omega \phi^{-2} = 8 \pi
G_{\rm eff}$.  The $k$-term is related to the total curvature of
the universe, and becomes negligible in the inflationary
epoch.  If we assume the following approximations for
the period of slow-roll,
%%%%%%%%[Equation 3]%%%%%%%%
\beqn
\nonumber \left| \frac{\dot{\phi}}{\phi} \right| \ll H ,\\
\left| \dot{\phi}^2 \right| \ll V(\phi) ,
\eeqn
and further assume that
%%%%%%%%[Equation 4]%%%%%%%%
\beq
\left| \frac{1}{a^2} \frac{\partial_i^2 \phi}{\phi} \right| \ll H^2 ,
\eeq
then the field equations may be approximated as follows
during the slow-roll period:
%%%%%%%%[Equation 5]%%%%%%%%
\beqn
\nonumber H^2 &\simeq& \frac{4 \omega}{3}
\frac{V(\phi)}{\phi^2} , \\
\ddot{\phi} + 3H \dot{\phi} - \frac{1}{a^2} \partial_i^2
\phi &\simeq& \frac{2\omega}{3 + 2\omega} \frac{\left(
4 V - \phi V^{\prime} \right)}{\phi} .
\eeqn
In equation (5), terms of order $(\partial_i \phi)^2$ have
been neglected.  We assume a generic Ginzburg-Landau
form for the (zero-temperature) potential,
%%%%%%%%[Equation 6]%%%%%%%%
\beq
V(\phi) = \frac{\lambda}{4} \left( \phi^2 - v^2
\right)^2 ,
\eeq
which describes a second-order phase transition (even
when non-zero-temperature corrections are included).
Following the ordinary procedure~\cite{guthpi82}, we
may write the field $\phi (\vec{\rm x}, t) = \varphi (t) +
\delta \phi (\vec{\rm x}, t)$, where $\varphi$ is the
homogenous part of the field, and $\delta \phi$
represents a quantum fluctuation of the field.  (This
justifies neglecting terms of order $(\partial_i \phi )^2$
above:  these terms are quadratic in the small
fluctuations.)  Then during slow-roll, we may further
ignore the $\ddot{\varphi}$ term, and solve the
approximate field equations (5) in terms of $\varphi (t)$:
%%%%%%%%[Equation 7]%%%%%%%%
\beqn
\nonumber \varphi (t) &=& \varphi_o +
\sqrt{\frac{\lambda \omega}{3 \gamma^2}} \> v^2 \> t ,
\\
\frac{a(t)}{a_B} &=& \left( \frac{\varphi (t)}{\varphi_o}
\right)^\gamma \exp \left[ \frac{\gamma}{2v^2} \left(
\varphi_o^2 - \varphi^2 (t) \right) \right] ,
\eeqn
where $\gamma \equiv  \omega + 3/2$.  The
quantities $\varphi_o$ and $a_B$ are values at the
beginning of the inflationary epoch.  For this Letter, we will be
concerned with a \lq new inflation'-like scenario, with $\varphi_o \ll
v$, instead of the \lq chaotic inflation' condition, $\varphi_o \gg v$.
Note that with $\varphi_o \ll v$, IgI inflates like a power law for early
times, $a(t) \sim t^\gamma$, much like the power-law expansion of
extended inflation. \\
\indent  If we define the quantity
%%%%%%%%[Equation 8]%%%%%%%%
\beq
W(\phi) \equiv \phi^{-1} \left[ \phi V^{\prime} - 4
V(\phi) \right] ,
\eeq
then the equation for the evolution of the fluctuations
may be written:
%%%%%%%%[Equation 9]%%%%%%%%
\beq
\delta \ddot{\phi} + 3H \delta \dot{\phi} + \frac{k^2}{a^2}
\delta \phi = - \frac{\omega}{\gamma} \left( W^{\prime}
|_{\varphi} \right) \delta \phi ,
\eeq
where $\delta \ddot{\phi} \equiv \partial^2 (\delta \phi) / \partial
t^2$.  For the particular form of the potential considered here,
this gives:
%%%%%%%%[Equation 10]%%%%%%%%
\beq
W^{\prime} |_{\varphi} \rightarrow \lambda v^2 \left( 1
+ \frac{v^2}{\varphi^2} \right) ,
\eeq
and thus
%%%%%%%%[Equation 11]%%%%%%%%
\beq
\delta \ddot{\phi} + 3H \delta \dot{\phi} + \frac{k^2}{a^2}
\delta \phi = - \frac{\lambda \omega}{\gamma} v^2
\left( 1 + \frac{v^2}{\varphi^2} \right) \delta \phi .
\eeq
Note that the $k^2 = \vec{k} \cdot \vec{k}$ term in
equations (9) and (11) should not be confused with the
curvature $k$ term in equation (2). \\
\indent  We need to calculate the two-point correlation
function for the fluctuations obeying equation (11), which
in turn will yield the prediction for $n_s$.  The two-point
correlation function for a $free$, massless scalar field in a
metric expanding as $a(t) \propto t^{\gamma}$ is by now
well known.~\cite{abwise84}~\cite{lythstew92}  Our task now is
to check whether or not the more complicated wave
equation for the fluctuations in IgI can be
well-approximated by the free wave equation, so that these
earlier results for the correlation function may be
incorporated.  In fact, the free wave equation is a good
approximation to equation (11), as can be seen by the
following.  Using equation (7), we may write:
%%%%%%%%[Equation 12]%%%%%%%%
\beq
\frac{\dot{a}}{a} = \gamma \frac{\dot{\varphi}}{\varphi}
\left[ 1 - \frac{\varphi^2}{v^2} \right] .
\eeq
We will be interested in the two-point correlation
function at the time ($t_{HC}$) of last horizon-crossing during
inflation of the density perturbations at scales which
interest us (from $10^6$ to $10^{10}$ lightyears).  The
time of last horizon-crossing is very difficult to solve for
exactly, but should have happened around 60 $e$-folds
before the end of inflation.  As in~\cite{DK94}, we may
write:
%%%%%%%%[Equation 13]%%%%%%%%
\beq
e^\alpha \equiv \frac{a (t_{end})}{a (t_{HC})} \sim e^{60} .
\eeq
Then equations (12) and (13) yield
%%%%%%%[Equation 14]%%%%%%%%%
\beq
\frac{\alpha}{\gamma} = \ln \left( \frac{v}{\varphi_{HC}}
\right) - \frac{1}{2} +
\frac{1}{2} \left( \frac{\varphi_{HC}}{v} \right)^2 .
\eeq
Following~\cite{AZT85}, we may study two limiting cases:
(a) $(\alpha / \gamma) \ll 1$ and (b) $(\alpha / \gamma) \gg 1$.
These give:
%%%%%%%[Equation 15]%%%%%%%%%
\beqn
\nonumber (\rm a)\>\>\> \it \varphi_{HC} \rm &\simeq& \it
v \rm \left( 1 - \sqrt{\frac{\alpha}{\gamma}} \right) , \\
(\rm b)\>\>\> \it \varphi_{HC} \rm &\simeq& \it v \rm
\exp \left(- \frac{\alpha}{\gamma} - \frac{1}{2} \right) .
\eeqn
The corresponding expressions in~\cite{AZT85} are written
incorrectly in terms of $\epsilon \equiv 1 / (4 \omega)$
instead of $\gamma$, because those authors made the
approximation that $\gamma \equiv \omega + 3/2 \rightarrow
\omega$ throughout their analysis.  As pointed out in~\cite{DK94}
{}~\cite{FakUn90}, this is an unnecessary restriction on $\omega$ which
can lead to qualitatively incorrect results. \\
\indent  We may check that each of these approximate values
for $\varphi_{HC}$ falls safely within the domain of the
slow-roll approximation by comparing $\varphi_{HC}$ with the
value of the field for which the slow-roll approximation
breaks down; that is,
the point at which $(\ddot{\varphi} + \dot{\varphi}^2
/ \varphi) = 3H \dot{\varphi}$, instead of being $\ll 3H
\dot{\varphi}$.  As calculated in~\cite{DK94}, this occurs
at
%%%%%%%%[Equation 16]%%%%%%%%
\beq
\varphi_{bd} = v \left[ 1 + \frac{1 - \sqrt{1 + 6
\gamma}}{3 \gamma} \right]^{1/2} .
\eeq
For $\gamma \gg \alpha$ (case (a)), this expression may be
written $\varphi_{bd} \simeq v (1 - \sqrt{1 / (6 \gamma)} )$, and
it is clear that $\varphi_{HC} < \varphi_{bd}$.  There is
a lower bound on $\varphi_{bd}$ which becomes relevant
for the case $\gamma \ll \alpha$ (case (b)): even when $\omega
\sim 0$, $\gamma \geq 3/2$, and $\varphi_{bd}
\geq 3v / (1 + \sqrt{10}) = 0.72\> v$. Yet for
case (b), $\varphi_{HC} \leq v\> \exp (-3/2) = 0.22\>v$. Thus
$\varphi_{HC} < \varphi_{bd}$ for both cases (a) and (b). \\
\indent Armed with these expressions for $\varphi_{HC}$,
we may return to equation (11) for $\delta \phi$:
%%%%%%%%[Equation 17]%%%%%%%%
\beq
\delta \ddot{\phi} + 3H \delta \dot{\phi} = -
\frac{\lambda \omega}{\gamma} v^2 \left( 1 +
\frac{v^2}{\varphi^2} \right) \delta \phi  - \frac{k^2}{a^2}
\delta \phi .
\eeq
We want solutions of this equation for times near
$t_{HC}$.  The coefficient of the
second term on the RHS at $t_{HC}$ is
%%%%%%%%[Equation 18]%%%%%%%%
\beq
\frac{k^2}{a^2 (t_{HC})} = H_{HC}^2 =
\frac{\lambda \omega}{3} \frac{ \left( v^2 -
\varphi_{HC}^2 \right)^2}{\varphi_{HC}^2} ,
\eeq
and thus the ratio of the two coefficients on the RHS of
equation (17) is
%%%%%%%%%[Equation 19]%%%%%%%%%
\beq
R \equiv \frac{k^2 / a^2 (t_{HC})}{\lambda
(\omega / \gamma) v^2 (1 + v^2 / \varphi^2_{HC})} =
\frac{\gamma}{3} \left[ \frac{1 - 2 (\varphi_{HC} / v)^2
+ (\varphi_{HC} / v)^4}{1 + (\varphi_{HC} / v)^2} \right] .
\eeq
For $\alpha / \gamma \ll 1$ (case (a)), this ratio becomes
%%%%%%%%%%%%[Equation 20]%%%%%%%%%%
\beq
\rm (a) \>\>\> \it R \rm = \frac{\gamma}{3}
\left( \frac{\alpha}{\gamma} \right)
\left( \frac{4 - 4\sqrt{\alpha / \gamma}}{2 -
2 \sqrt{\alpha / \gamma}}\right) = \frac{2 \alpha}{3} \gg 1 .
\eeq
Thus, for case (a), the second term on the RHS of equation (17)
dominates near $t_{HC}$, and the equation for the fluctuations
assumes the form for a free, massless scalar field:
%%%%%%%%%%[Equation 21]%%%%%%%%%
\beq
\delta \ddot{\phi} + 3H\delta \dot{\phi} + \frac{k^2}{a^2}
\delta \phi \simeq 0 .
\eeq
For $\alpha / \gamma \gg 1$ (case (b)), $R$ becomes
%%%%%%%%%%%%%%%%[Equation 22]%%%%%%%%
\beq
\rm (b) \>\>\> \it R = \rm  \frac{\gamma}{3} \left(
\frac{1 - 2\exp(-2\alpha/\gamma - 1) + \exp(-4\alpha/
\gamma - 2)}{1 + \exp(-2\alpha/\gamma - 1)} \right) \simeq
\frac{\gamma}{3} ,
\eeq
and the fluctuations obey the equation
%%%%%%%%%%%%%[Equation 23]%%%%%%%%
\beq
\delta \ddot{\phi} + 3H\delta \dot{\phi} +
\left(1 + \frac{3}{\gamma} \right) \frac{k^2}{a^2}
\delta \phi \simeq 0 .
\eeq
Note that $(1 + 3 / \gamma) \leq 3$ for all $\gamma$.  (This
is an example of how the corrections to~\cite{AZT85} can
become crucial:  even when $\omega$ is made arbitrarily
small, the ratio $R^{-1}$ remains finite.)  The deviation of
equation (23) from the truly free, massless case is
thus small, and, furthermore, does not alter the
$k$-dependence of the two-point correlation function
(although it does affect the magnitude of the
correlation function).  Thus, for
both cases (a) and (b), we may import the results
from~\cite{abwise84}~\cite{lythstew92}:  writing the correlation
function as
%%%%%%%%[Equation 24]%%%%%%%%
\beq
\left| \Delta \phi (\vec{k}) \right|^2 \equiv k^3 \int
\frac{d^3 x}{(2 \pi)^3} e^{i \vec{k} \cdot \vec{x}} \langle
\phi (\vec{x}) \phi (\vec{0}) \rangle
\eeq
leads to the result
%%%%%%%%[Equation 25]%%%%%%%%
\beq
\left| \Delta \phi (\vec{k}) \right|^2 \propto k^{-2 /
(\gamma - 1)} .
\eeq
Therefore $\delta_H = \delta \tilde{\rho} / \rho \propto k^{-1 /
(\gamma - 1)}$.  The spectral index is defined by~\cite{pr93}
%%%%%%%%[Equation 26]%%%%%%%%
\beq
n_s - 1 \equiv \frac{d \ln \delta_H^2}{d \ln k} ,
\eeq
which yields
%%%%%%%%[Equation 27]%%%%%%%%
\beq
n_s = 1 - \frac{2}{\gamma - 1}
\eeq
for IgI.  As calculated in~\cite{DK94}, values of $\omega$ in the
range $10^2 \leq \omega \leq 10^3$ are favored for IgI,
based on the upper bound on the quartic self-coupling
parameter, $\lambda$.  (See Figure 2 in~\cite{DK94}.)  In particular, the
bound on $\lambda$ is maximized for $\omega_{cr} = 240$.
Equation (27) yields $n_s(\omega_{cr}) = 0.99$; for
$10^2 \leq \omega \leq 10^3$, IgI predicts $0.98 \leq n_s
\leq 1.00$.  This is obviously
quite close to the $n_s = 1.00$ Harrison-Zel'dovich spectrum. \\
\indent  Recently, the results of the two year
data analysis for the COBE Differential Microwave Radiometer (DMR)
experiments were announced.~\cite{gorski94}  The maximum likelihood
estimates on $n_s$ were given as 1.22 if the quadrupole
contribution were included, and 1.02 if the quadrupole
were excluded.  The marginal likelihood gave $n_s = 1.10
\pm 0.32$ including the quadrupole, and $n_s = 0.87 \pm
0.36$ excluding the quadrupole.  As concluded
in~\cite{gorski94}, these results are completely consistent
with an $n_s = 1.00$ spectrum, and hence are in close
agreement with the predictions from IgI.  Furthermore, it is
clear that these data imply a lower limit on $\omega$ for IgI,
based on equation (27).  In particular, if we
limit $n_s \geq 0.87$, then $\omega \geq 15$. \\
\indent  It should be noted, however, that IgI appears to
be incapable of yielding a spectrum with $n_s > 1.00$.
This could lead to conflict with the experimental value, if
$n_s$ should be shown definitively to hover closer to
$1.2$, rather than to $1.0$.  This illustrates how the
$spectrum$ of perturbations functions as a test for
models of the early universe, independently from the test
based on the $magnitude$ of perturbations:  it is always
possible (even if aesthetically unappealing) to rescale
dimensionless constants of inflationary models in order to match
the experimentally-determined magnitude of
perturbations.  Yet no such rescaling (at least in IgI) can
lead to a prediction of $n_s > 1$.  Inflationary models
which do predict $n_s \geq 1$ are considered
in~\cite{moll93}~\cite{schaefer94}. \\
\indent  The lower bound on $\omega$, stemming from
requirements on $n_s$, may prove to be very significant when treating the
magnitude of density
perturbations in IgI. In~\cite{FakUn90}, Redouane Fakir and William
Unruh proposed that constraints
on the magnitude of perturbations could be met in a
\lq chaotic inflation' scenario ($\varphi_o \gg v$) of IgI if
$\xi  \equiv 1/ (4 \omega)$ were
made arbitrarily large.  As the above analysis has shown,
however, in the context of a \lq new inflation' scenario ($\varphi_o \ll
v$), $\xi$ $cannot$ be made arbitrarily large: as
$\omega \rightarrow 0$, $\gamma \rightarrow 3/2$, and $n_s \rightarrow
-3$, which is in clear violation of the experimental data.   The
differing constraints from $n_s$ on $\xi$ (or, equivalently, on $\omega$)
for the \lq chaotic inflation' versus the \lq new inflation' scenarios
are further studied in~\cite{dkprd}. \\
\indent  Finally, we must consider why
IgI is able to agree with the experimental determination of $n_s$, even
though the closely-related extended inflation cannot.  The
result of equation (27) is similar in form to that for
extended inflation, for which $\gamma$ in equation (27)
should be replaced by $\gamma / 2$.  Yet, as explained
in~\cite{lidlyth92}, extended inflation is restricted to
$\omega \leq 17$, which leads to $n_s \sim 0.76$.  This
is too steep a tilt away from the nearly scale-invariant,
$n_s = 1.00$ spectrum, and is thus a problem for that
model.  The restriction on $\omega$ for extended
inflation arises from that model's first order phase
transition:  values of $\omega$ greater than 17 would
yield noticeable (and yet unseen) inhomogeneities in the
cosmic microwave backround radiation, due to the
percolation of such a large range in bubble sizes.  This has
often been referred to as the \lq big bubble' or \lq
$\omega$ problem'.~\cite{lidlyth92}~\cite{bubble}  In IgI, the
second order phase transition removes concern with
bubble percolation, and there is no analogous upper
bound on $\omega$. \\
\indent  The second \lq $\omega$ problem' for extended
inflation, which is also avoided in IgI, concerns the
present-day value of the Brans-Dicke parameter
$\omega$.  Time-delay tests of Brans-Dicke gravitation
versus Einsteinian general relativity limit $\omega \geq
500$.  Since extended inflation never exits the GET phase,
this present-day bound on $\omega$ should be obeyed,
which is in further violation of the $\omega \leq 17$
bound.  In IgI, the phase transition itself ensures the
transition from the GET phase to pure
Einsteinian general relativity; that is, after the very
early phase transition,
the universe is no longer described by Brans-Dicke
gravitation, and all present-day bounds on $\omega$
become irrelevant.  IgI delivers the universe into the
highly-corroborated Einsteinian gravity, regardless of the
value of $\omega$ during the early universe. In addition, as discussed
in~\cite{DK94}, post-inflation reheating for IgI could bring the universe
to energies as high as $(10^{-3} - 10^{-2})M_{pl}$, where $M_{pl} = 1.22
\times 10^{19}$ GeV is the present value of the Planck mass, allowing
$either$ GUT-scale or electroweak baryogenesis~\cite{mg} to follow the IgI
phase.\\
\indent  Thus, Induced-gravity Inflation predicts a
spectrum of primordial density perturbations in close
agreement with the recent COBE DMR results, with
$0.98 \leq n_s \leq 1.00$.  Using the experimental data, we
may limit $\omega \geq 15$ for Induced-gravity Inflation.
Although the functional form of the spectral index predicted by
Induced-gravity Inflation is very similar to that for
extended inflation, Induced-gravity Inflation is able to
avoid $both$ of the \lq $\omega$ problems' which plagued
extended inflation.  The predicted spectrum, therefore,
deviates little from the observed, nearly
scale-invariant spectrum. \\
\section{Acknowledgments}
It is a pleasure to thank Eric Carlson for many helpful
discussions during the course of this work, and Redouane Fakir for
supplying a copy of ref.~\cite{FakHabib}.
This research was supported by an NSF Fellowship for Predoctoral Fellows,
and partly by NSF Grant PHY-92-18167. \\
%
%------------------------------------------------------

%
%
\end{document}